# The electronic structure and dipole moment of charybdotoxin, a scorpion venom peptide with $K^+$ channel blocking activity

**Fabio Pichierri***

*G-COE Laboratory, Department of Applied Chemistry, Graduate School of Engineering, Tohoku University, Aoba-yama 6-6-07, Sendai 980-8579, Japan*

## Abstract

The electronic structure of charybdotoxin (ChTX), a scorpion venom peptide that is known to act as a potassium channel blocker, is investigated with the aid of quantum mechanical calculations. The dipole moment vector (μ=145 D) of ChTX can be stirred by the full length KcsA potassium channel's macrodipole (μ=403 D) thereby assuming the proper orientation before binding the ion channel on the cell surface. The localization of the frontier orbitals of ChTX has been revealed for the first time. HOMO is localized on Trp14 while the three lowest-energy MOs (LUMO, LUMO+1, and LUMO+2) are localized on the three disulfide bonds that characterize this pepetide. An effective way to engineer the HOMO-LUMO (H-L) gap of ChTX is that of replacing its Trp14 residue with Ala14 whereas deletion of the LUMO-associated disulfide bond with the insertion of a pair of L-α-aminobutyric acid residues does not affect the H-L energy gap.

*Keywords:* Charybdotoxin; Scorpion venom peptide; Potassium channel; Dipole moment; Electronic structure; Quantum chemistry

* Corresponding author. Tel. & Fax: +81-22-795-4132

*E-mail address:* fabio@che.tohoku.ac.jp (F. Pichierri)

## 1. Introduction

The venom of scorpions contains a pool of several globular peptides (mini-proteins) which have the ability to bind a variety of ion ($Na^+$, $K^+$, $Ca^{2+}$) channels located on the cell surface of the organism under attack [1-3]. Upon binding to the extracellular side of the ion channel's pore, the ions' flux through the membrane is blocked and this event triggers a cytokine-mediated inflammatory response which produces local pain and can escalate up to respiratory failure and death [4]. In 1982 the first scorpion venom peptide, noxiustoxin, was isolated from the venom of the scorpion *Centruroides noxius* [5]. Soon after, many other scorpion toxins were identified and structurally characterized (mainly by solution NMR spectroscopy). To date the 3D structures of more than one hundred scorpion venom peptides have been characterized and their atomic coordinates deposited into the Protein Data Bank (PDB) [6]. Beside their toxicity, these peptides might find useful applications as drugs in the treatment of immunological diseases [7,8] and as insecticides [9].

Among the venom peptides investigated so far, the 37 amino acid (AA) residue charaybdotoxin (ChTX) has played a special role as channel blocker. ChTX was discovered by Miller et al. [10] in 1985 after its isolation from the venom of scorpion *Leiurus quinquestriatus*. The molecular structure of ChTX was subsequently characterized by Bontems et al. [11,12] using NMR spectroscopy. As shown in Figure 1(a), ChTX is made of a small α-helix (containing 10 AA residues) linked to a three-stranded antiparallel β-sheet via a pair of disulfide bonds (cystine moieties). A third disulfide bond connects the β-sheet to a loop that protrudes from the α-helix. This characteristic α/β motif is widely conserved among scorpion toxins. The N-terminal residue of ChTX is the neutral pyroglutamic acid (PGA) or pyrrolidone carboxylic acid (denoted to as Z in Figure 1(b)), an uncommon amino acid which is known as being an N-terminal blocker that arises from the cyclization of N-terminal glutamine residues [13]. The C-terminal residue of ChTX is a serine (Ser37) which is linked via its $C_{\alpha}$ atom to the negatively-charged carboxylate group.

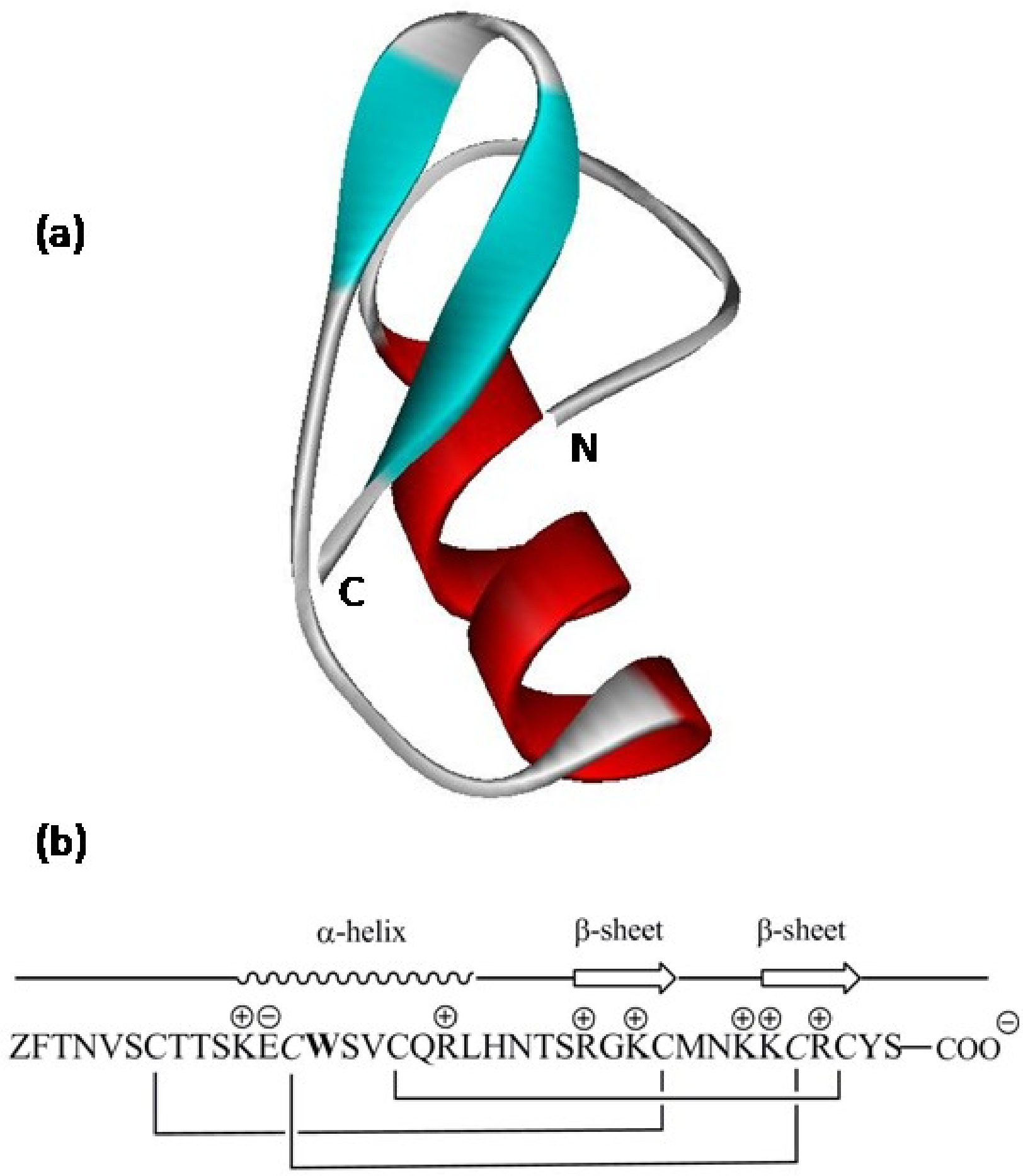

**Figure 1.** (a) Ribbon representation of the 3D structure of ChTX (standard orientation). (b) Primary structure of ChTX with the corresponding elements of secondary structure and disulfide bonds between cytosine (C) residues. The localization of the HOMO on Trp14 (**W**) and LUMO on Cys13-Cys35 (*C*) residues is also shown along with the formal charge on each amino acid side chain.

In 2005, almost twenty years after the discovery of ChTX, Yu et al. [14] determined the structure of the complex between ChTX and KcsA potassium channel by using NMR spectroscopy. The authors introduced six mutations into KcsA so as to increase its affinity toward ChTX. The experimental structure of the complex (PDB id 2A9H) evidenced for the first time the formation of important contacts among the AA residues of these two proteins. In particular, it emerged that a charged lysine residue (K27) located on the β-sheet of ChTX is found being inserted into the ion channel's selectivity filter thereby blocking the passage of $K^+$ ions. Although no other experimental structure of venom peptide/channel complex is known, the 3D models of the complexes between ChTX and other potassium channels (Kc1.2, BK) have been investigated computationally using a combination of homology modeling, docking, and molecular dynamics (MD) simulation [15-17].

As far as the molecular recognition of K channels is concerned, the permanent dipole moment of these peptides has been suggested to play an important role in determining the orientation of the peptide when it approaches its target ion channel on the cell surface [16]. In this regard, a recent MD simulation study by Chen and Kuyukak [17] indicated that upon unbinding of ChTX from KcsA the dipole moment vector of the peptide aligns with the dipole vector of the ion channel whereas in the ChTX:KcsA complex, whereby strong electrostatic interactions are established at the interface, the two dipole vectors form an angle of 40°.

In this study we investigate the electronic structure of ChTX using quantum mechanical calculations [18]. Along with the frontier orbitals and distribution of electronic charge in each amino acid side chain, we wish to provide a quantitative estimate of the magnitude and orientation of the permanent dipole moment of this important peptide. So far, and to the best of our knowledge, the magnitude of the dipole moment of ChTX is as yet unknown. Using classical force fields Yu et al. [15] showed that the dipole vector of ChTX and those of five other venomous peptides are oriented toward their β-sheet regions in the corresponding molecular structures but the magnitude of each dipole moment was not given. On the other hand, Ireta et al. [19] performed total energy pseudopotential calculations to investigate the electronic structure of ChTX using the experimental NMR structure of Bontems et al. [11,12]. With the aid of a band structure approach they obtained useful information on the electrostatic potential and chemical softness but the dipole moment was not investigated. It is therefore timely to quantitatively characterize the dipole moment of ChTX also in the light of recent theoretical results of the author concerned with the dipole moment of KcsA potassium channel [20].

## 2. Computational details

The solution NMR structure of ChTX determined by Bontems et al. [11,12] and whose atomic coordinates are stored in the PDB (accession number 2CRD) was employed here. Of the 12 model conformers that are included in the PDB file, the first one was selected. Hydrogen atoms were same as those in the experimental structure. Quantum mechanical calculations were performed with the MOPAC2009 software package [21] of Stewart which contains the MOZYME module [22] and novel PM6 parameters [23] for the QM treatment of proteins [24]. First, we determined the net charge of ChTX by computing its Lewis structure which is made of 585 σ-type bonds, 179 lone-pairs, and 55 π-type bonds. There are seven positive charges and two negative charges which contribute a total net charge of +5. Hence, the net charge is same as that of the ChTX model of Chen and Kuyukak [17]. The resulting atomic model of ChTX contains 577 atoms (empirical formula: $C_{176}H_{282}N_{57}O_{55}S_7$) and 41 peptide linkages. Next we optimized the geometry of this protein. Because ChTX is a small peptide (37 amino acid residues), the canonical molecular orbital (MO) optimization routines of MOPAC2009 were employed in combination with the COSMO solvation model of Schüürmann [25] (for water the recommended value of ε=78.4 for the dielectric constant was employed). The level of theory employed in this study is thus denoted to as PM6-SCF-MO-COSMO. The JMol software [26] was employed for the visualization of MOs while the dipole moment vector was visualized with a local software.

## 3. Results and discussion

### *3.1 Geometry*

The geometry of ChTX optimized at the PM6-SCF-MO-COSMO level of theory is shown in Figure 2. On the left side of Figure 2 is depicted the optimized structure of CT in the standard orientation (see Figure 1) while on the right side of Figure 2 the structure has been rotated by 90° about the vertical axis passing through Lys31. As seen from this figure, a number of positively charged amino acid residues are sticking out from the protein, namely Lys31 and Lys 32 on the top, Lys27 and Arg34 in the middle, and Lys11, Arg19, and Arg25 at the bottom. Among these charged amino acids, Lys27 forms an H-bond with the oxygen atom of Tyr36 and the nitrogen atom of Arg34. The latter group interacts also with the oxygen lone-pair of Tyr36 thereby contributing to the stabilization of this Lys27/Arg34/Tyr36 triad, as depicted in Figure 3(a). Arg25 forms a salt bridge with the C-terminal carboxylate group where Ser37 is located, as displayed in Figure 3(b). Hence, it appears that the remaining lysine and arginine chains are likely being involved in interactions with the solvent (here assumed as a continuum).

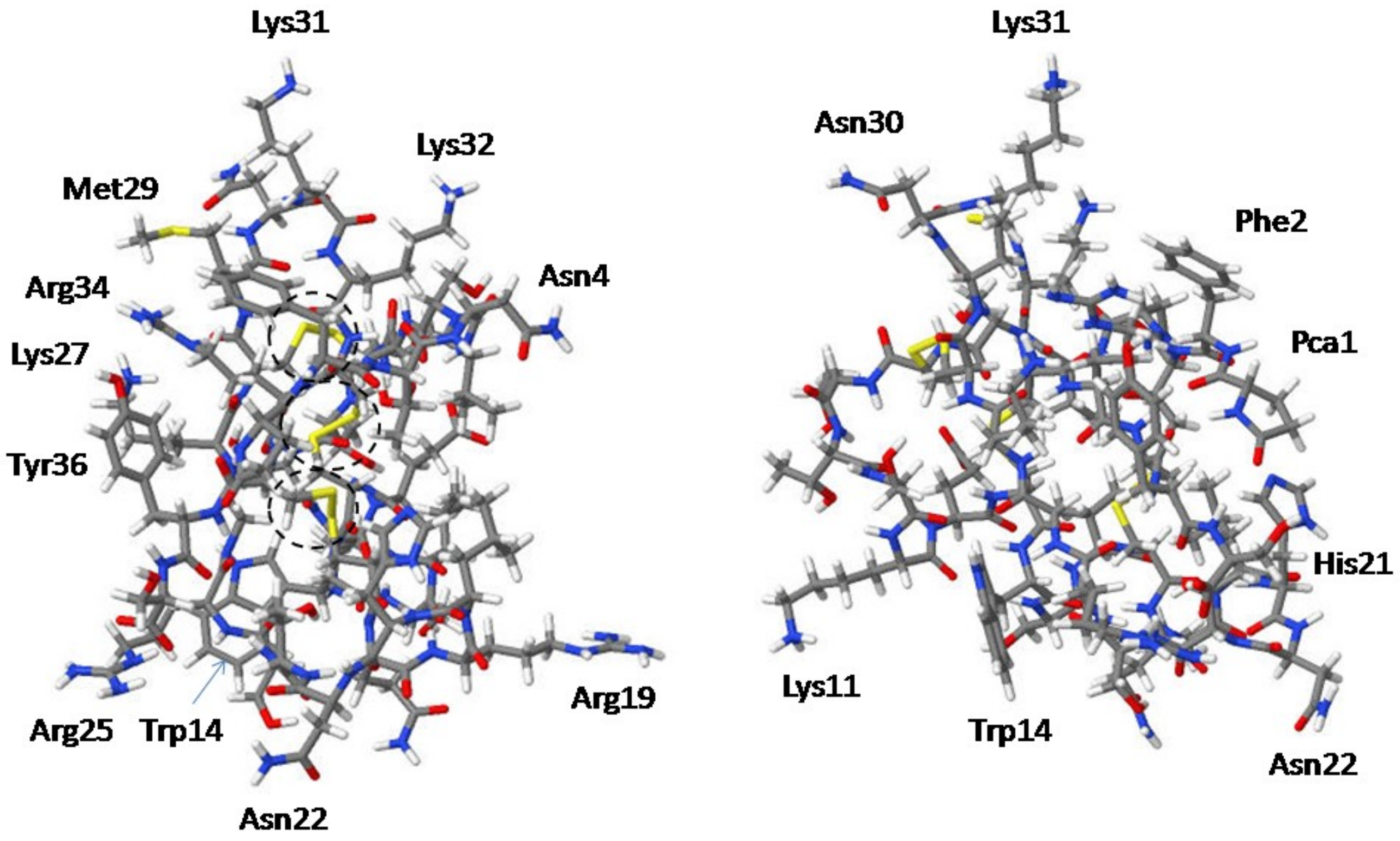

**Figure 2.** Optimized geometry of ChTX shown in the standard orientation (left) and after a 90° rotation about the axis passing through Lys31. Disulfide bonds (cystine moieties) are circled (left) and selected amino acids chains are labeled.

Several intramolecular H-bonds of type C=O···H–N are established among peptide units in the α-helix and β-sheet elements of ChTX. In addition, the carbonyl and NH groups of two consecutive peptide linkages form H-bonds with the NH moiety of Trp14, as shown in Figure 3(c). These H-bonds are likely to stabilize the orientation of Trp14 on the protein surface. Furthermore, the carbonyl group of a peptide linkage forms an H-bond with the N-terminal PGA residue, as shown in Figure 3(d). Of the three disulfide bonds, highlighted by circles in Figure 2, only one (Cys7-Cys28) is exposed to the solvent whereas the other two (Cys13-Cys33 and Cys17-Cys35) are buried inside the protein. The average S–S bond length computed here at the SCF-MO-PM6-COSMO level of theory corresponds to ~2.05 Å which is in excellent agreement with the experimental value of 2.0472(4) Å determined by an x-ray crystallographic analysis of L-cystine at 110 K [27]. It is worth noticing that the placement of the three disulfide bonds in the central part of the protein confers substantial rigidity to the whole tertiary structure of ChTX.

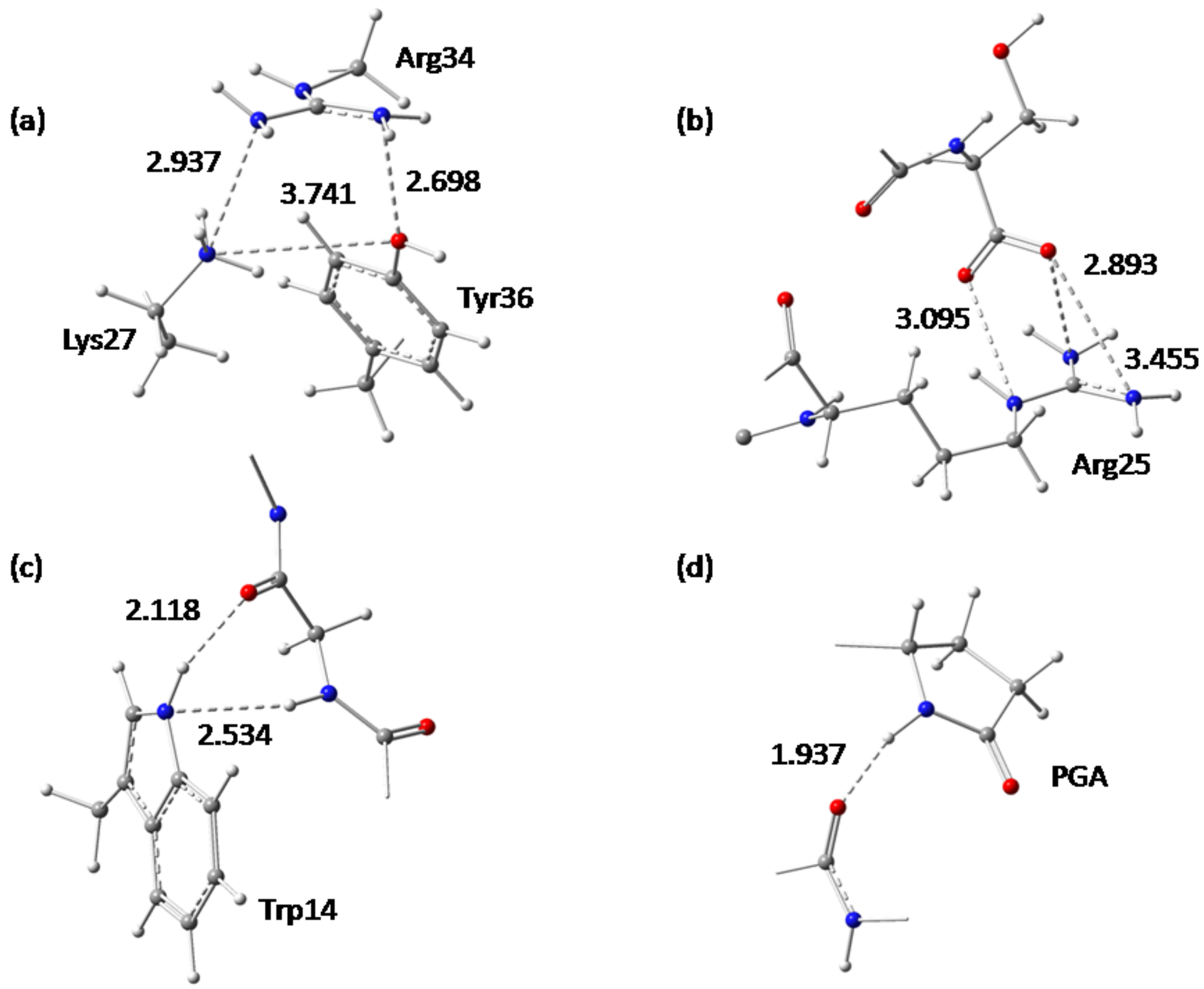

**Figure 3.** Details of the intramolecular contacts in ChTX: (a) the Lys27/Arg34/Tyr36 triad and distances between heteroatom pairs; (b) salt bridge between Arg25 and the C-terminal carboxylate group; (c) H-bonds between Trp14 and a nearby pair of peptide linkages; and (d) H-bond between PGA and a nearby peptide linkage. Distances in Å.

*3.2 Electronic structure*

We analyzed the frontier orbitals of ChTX which are depicted in Figure 4. The two highest-occupied molecular orbitals, HOMO and HOMO-1, are localized on the Trp14 residue while HOMO-2 is localized on one disulfide bond. On the other hand, the lowest-unoccupied molecular orbitals, LUMO, LUMO+1, and LUMO+2, are localized on the disulfide bonds of Cys13-Cys33, Cys7-Cys28, and Cys17-Cys35, respectively, as shown in Figure 4.

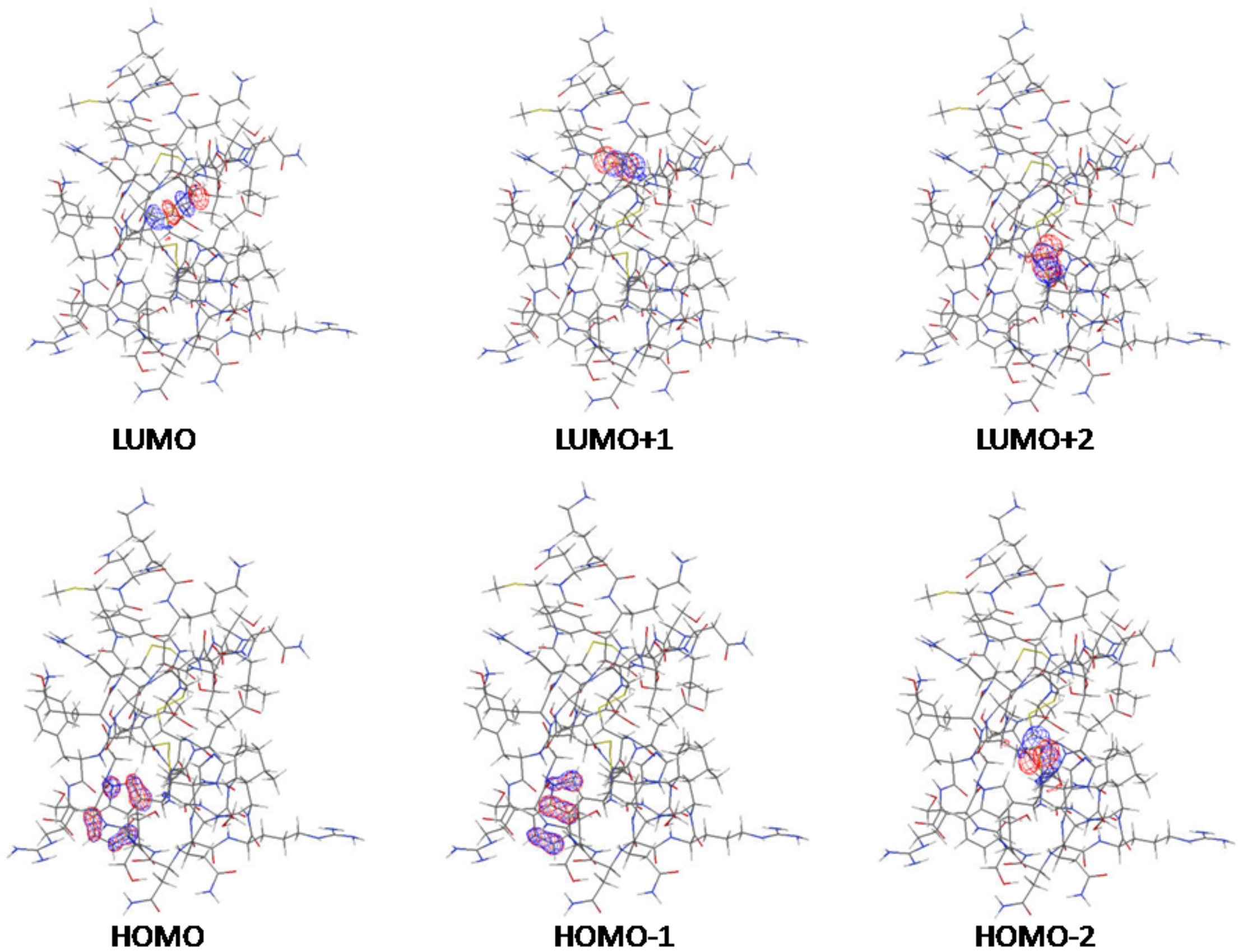

**Figure 4.** Frontier molecular orbitals of ChTX (standard orientation).

Interestingly, an analogous localization pattern of the frontier orbitals was found also in a recent quantum mechanical study of human erythropoietin (*hu*EPO) [28] where HOMO and HOMO-1 are localized on two tryptophan residues (Trp51 and Trp64) while LUMO and LUMO+1 are localized on two disulfide bonds (Cys7-Cys161 and Cys29-Cys33). Further, also the HOMO of the SH2 domain of p56lck tyrosine kinase is localized on a tryptophan residue (Trp127) [29]. Hence, removal of electrons from these proteins should occur through the oxidation of their tryptophan residues while the addition of electrons would involve the reduction of their disulfide bonds. Among the three disulfide bonds of ChTX, we notice from Figure 4 that Cys13-Cys33, where the LUMO is localized, is buried into the protein while Cys7-Cys28 bearing LUMO+1 is exposed to the solvent and hence it is the one more likely being subject to chemical reduction.

The computed HOMO-LUMO (H-L) gap of ChTX corresponds to 6.676 eV. This is slightly larger than the H-L gap of 5.534 eV computed (at the PM5-COSMO level) for the native structure of *hu*EPO [28] but smaller than the H-L gap of ubiquitin which corresponds to 8.346 eV [30]. We have shown that it is possible to modify the H-L gap of *hu*EPO by mutation of selected amino acid residues [28]. Another possibility for engineering the H-L gap is that of deleting disulfide bonds. Both strategies are feasible in the case of ChTX as a number of mutants have been investigated by Miller's group [31,32] while Song et al. [33] have been able to delete one disulfide bond (Cys13-Cys33) by inserting two L-α-aminobutyric acid residues (Cys13-Aba and Cys33-Aba). The structure of the resulting two-disulfide derivative of ChTX, termed CHABII, was determined by solution NMR spectroscopy and the atomic coordinates of 30 conformers deposited in PDB (code 1BAH). We have fully optimized the geometry of the first conformer (denoted to as model 1 inside the PDB file) at the SCF-MO-PM6-COSMO level of theory and the computed H-L gap corresponds to 6.666 eV, which is only 0.01 eV smaller than that of the native peptide. Because the three lowest-energy MOs of ChTX (localized on the three disulfide bonds) are very close in energy, the deletion of one disulfide bond is not effective in changing the H-L gap of the peptide. A more effective way of engineering the H-L gap of ChTX would be that of replacing the Trp14 residue, where HOMO is localized, with a non-aromatic residue such as alanine. We therefore computed the electronic structure of the Trp14Ala mutant of ChTX and obtained an H-L gap of 7.266 eV, which is 0.59 eV larger than that of the native peptide. Hence, replacing the residue where HOMO is localized represents an effective strategy for engineering the H-L gap of this protein.

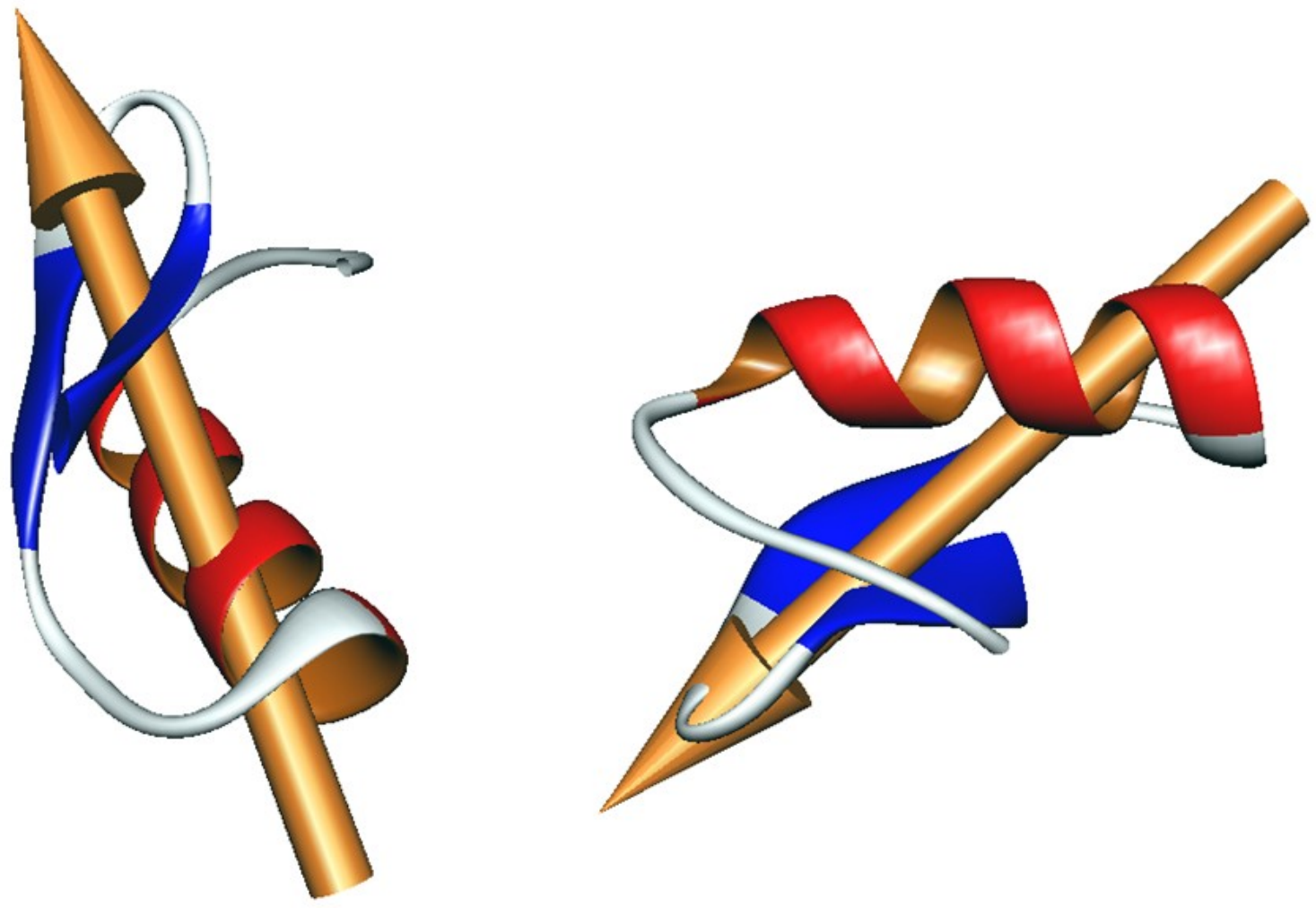

**Figure 5.** Dipole moment vector of ChTX with the peptide in the standard orientation (left) and oriented with the α-helix above the β-sheet. The arrow's center is placed on the center of mass of the protein.

### *3.3 Dipole moment*

The dipole moment of ChTX, which has a magnitude of 145 D, was computed with respect to the center of mass of the protein [21]. Figure 5 shows the dipole moment vector of ChTX as obtained at the PM6-SCF-MO-COSMO level of theory. As seen from this figure, the dipole vector forms an angle of ~50° with the α-helix while the positive tip is oriented towards the β-sheet. It has been suggested that when ChTX approaches a K channel such as KcsA, the peptide will align its dipole moment vector with respect to the dipole of the channel [17]. Accordingly, quantum mechanical calculations recently performed by the author indicated that the truncated form of KcsA (TF-KcsA) has a dipole moment of 272 D while the full length (FL-KcsA) structure possesses a dipole of magnitude 403 D [20]. In both cases the dipole moment vector is aligned along the fourfold axis of the ion channel and with the positive head pointing toward the internal part of the cell while the negative end is located on the extracellular region. These results indicate that the macrodipole of Kcsa is large enough to provoke the alignment of ChTX dipole thereby assisting the scorpion venom peptide in positioning its β-sheet toward the channel's mouth, as seen in the experimental structure depicted in Figure 6.

Upon binding, however, stronger electrostatic interactions are formed at the ChTX/channel interface which provoke the bending of ChTX thereby changing the relative orientation of the two dipoles. In this regard, Chen and Kuyukak [17] found that the dipole vectors of the two proteins form an angle of 40° once they are tightly bonded in the complex. By placing the structure of ChTX above the KcsA mouth and adjusting its orientation according to that in the experimental structure of the complex, we were able to determine an angle of ~50° between the dipole moments of the two proteins. This result is in excellent agreement with that of Chen and Kuyukak [17] especially if one takes into account that the formation of the ChTX:KcsA complex will modify the orientation of the amino acid chains that are located at the interface with respect to the free peptide. Also, one has to consider that the large dipole moments of TF-KcsA (μ=272 D) and FL-KcsA (μ=403 D) will polarize the electronic charge density of ChTX thereby optimizing the interaction of these two molecules in the complex.

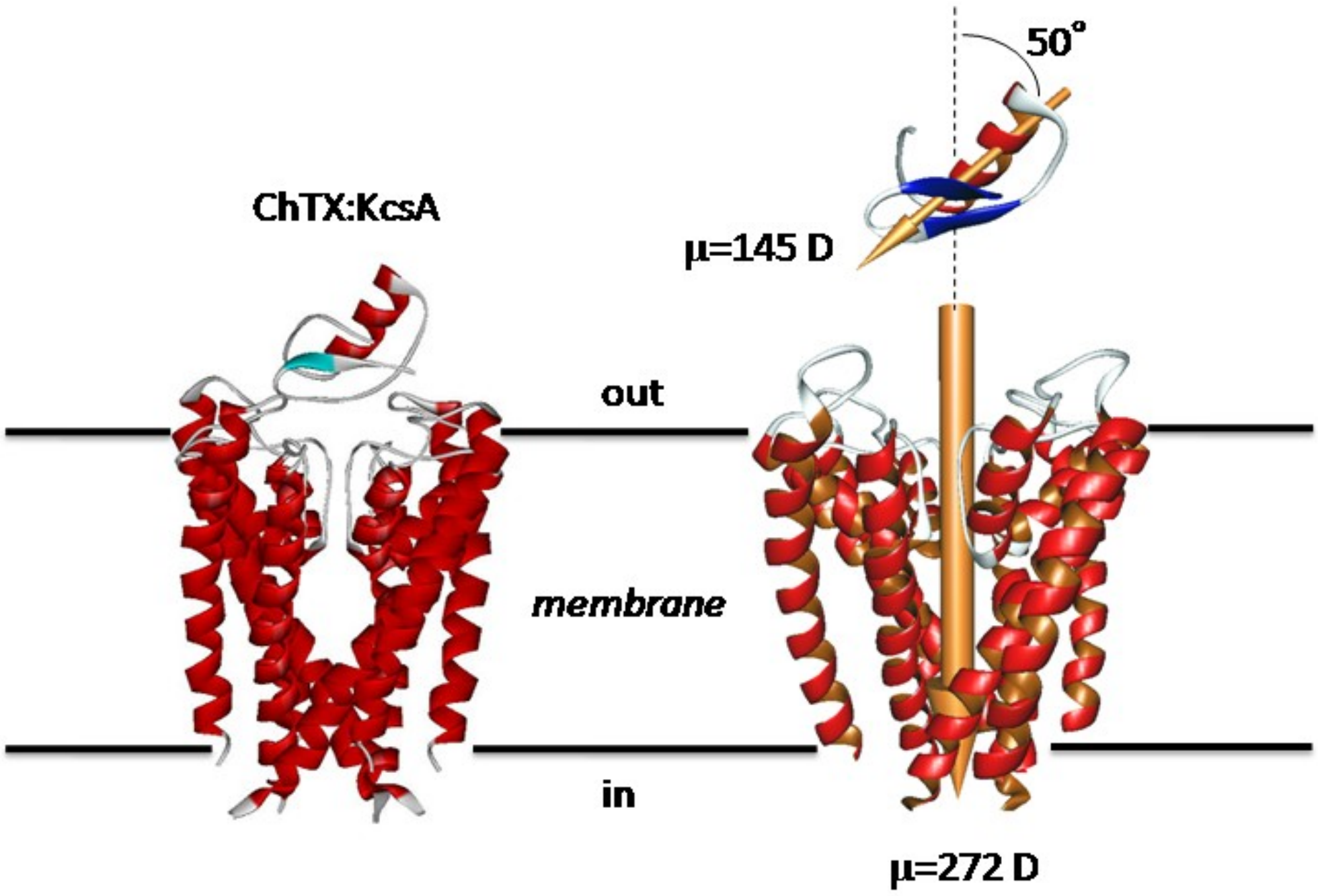

**Figure 6.** Comparison between the dipole moments of ChTX and KcsA potassium channel (right). ChTX is oriented as in the experimental structure of the ChTX:KcsA complex (left) but moved away from the channel's mouth.

## 4. Conclusions

We investigated the electronic structure of ChTX, a scorpion venom peptide which is known to act as a potassium channel blocker, with the aid of quantum mechanical calculations. The results indicate that the dipole moment vector of ChTX is oriented toward the β-sheet while forming an angle of ~50° with the α-helix. The magnitude of the dipole moment (μ=145 D) is such that it can be stirred by the macrodipole of KcsA thereby properly orienting the peptide before binding. Furthermore, the location of the frontier orbitals of ChTX has been revealed for the first time and an effective strategy for engineering the HOMO-LUMO gap of this scorpion peptide has been proposed.

## Acknowledgments

I thank Dr. James J.P. Stewart for a copy of MOPAC2009 and for useful advice on the proper usage of the software. Stimulating discussions with Prof. Serdar Kuyucak are gratefully acknowledged. This work is supported by the Global COE program (IREMC) and the Graduate School of Engineering of Tohoku University. This work is funded by Grant-in-Aid for Scientific Research [Kakenhi C 20550146].